\documentclass[twocolumn,nofootinbib]{revtex4}
\input psfig.sty
\usepackage[dvips]{color}
\usepackage{graphicx}
\usepackage{epsfig}

\begin{document}

\title{Inflation driven by particle creation}

\author{Saulo Carneiro\footnote{E-mail: saulo.carneiro@pq.cnpq.br.}}

\affiliation{Instituto de F\'{\i}sica, Universidade Federal da Bahia, Salvador, BA, Brazil\\International Centre for Theoretical Physics, Trieste, Italy\footnote{Associate member.}}

\date{\today}

\begin{abstract}
The creation of ultra-light dark particles in the late-time FLRW spacetime provides a cosmological model in accordance with precise observational tests. The matter creation backreaction implies in this context a vacuum energy density scaling linearly with the Hubble parameter $H$, which is consistent with the vacuum expectation value of the QCD condensate in a low-energy expanding spacetime. Both the cosmological constant and coincidence problems are alleviated in this scenario. We also explore the opposite, high energy limit of the particle creation process. We show that it leads to a non-singular primordial universe where an early inflationary era takes place, with natural reheating and exit. The generated primordial spectrum is scale invariant and, by supposing that inflation lasts for $60$ e-folds, we obtain a scalar expectral index $n \approx 0.97$.
\end{abstract}

\maketitle

\section{Introduction}

The gravitational role of vacuum fluctuations is a challenging problem in field theory and cosmology, as vacuum is in general not uniquely defined in curved spacetimes and its energy density usually depends on the renormalization procedure. In the case of free (conformal) fields in de Sitter spacetime, the renormalized vacuum density is $\Lambda \approx H^4$ \cite{H4}. In a low-energy universe this leads to a too tiny cosmological term. In the high energy limit it can be used to obtain a non-singular model, with an initial quasi-de Sitter universe giving origin to a radiation era, through a phase transition in which vacuum decays into relativistic particles \cite{2006}. The primordial perturbations generated in this process are not scale invariant and a subsequent inflation is still needed \cite{Tavakol}.

In order to see how this non-singular scenario emerges, let us take $p_{\Lambda} = - \Lambda$, which is valid in de Sitter space-time owing to its maximal symmetry. The conservation equation
\begin{equation} \label{conservation}
\dot{\rho} + 3H(\rho + p) = 0
\end{equation}
assumes the form
\begin{equation} \label{noconservation}
\dot{\rho}_m + 3H(\rho_m + p_m) = - \dot{\Lambda},
\end{equation}
where $\rho = \rho_m + \Lambda$ and $p = p_m - \Lambda$ are the total energy and pressure, respectively, and the dot means derivative w.r.t. the cosmological time $t$. We see that a time-varying vacuum density is concomitant with particle production. 

Let us now take $\Lambda = 3 H^4$ and the relativistic equation of state $p_m = \rho_m/3$ for matter. Eq. (\ref{noconservation}) and the Friedmann equation
\begin{equation}
\rho_m + \Lambda = 3H^2
\end{equation}
lead to
\begin{equation}
\dot{H} + 2H^2-2H^4 = 0.
\end{equation}
Apart from the trivial solution $H = 1$, this equation has the solution
\begin{equation} \label{H4}
2t = \frac{1}{H} - \tanh^{-1}H,
\end{equation}
where a constant of integration was conveniently chosen. This solution, which tends to de Sitter in the asymptotic past, is plotted in Fig. 1. We then see that de Sitter solution is not stable in the high energy limit, evolving to a radiation dominated universe, with vacuum decaying into relativistic matter (it is easy to see from (\ref{H4}) that, when $H \rightarrow 0$, we have $Ht = 1/2$, which characterises a radiation phase). As we will see, this process involves a backreaction that eventually changes the solution (\ref{H4}), leading to a subsequent inflation era.

\begin{center}
\begin{figure}[t!]
\includegraphics[]{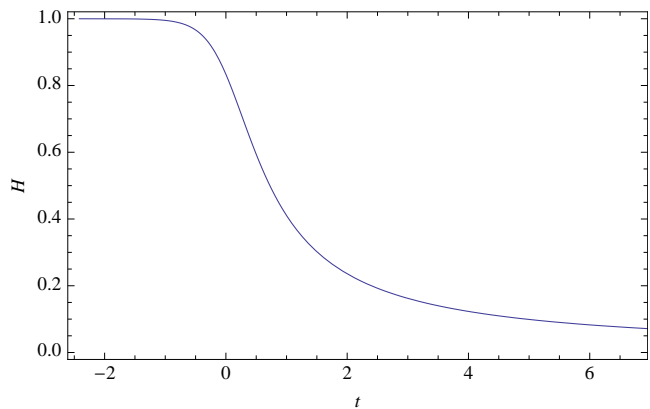}
\caption{{\protect\footnotesize The Hubble parameter as a function of time for $\Lambda=3H^4$.}}
\end{figure}
\end{center}

To clarify these results, let us revisit the seminal work by Gibbons and Hawking \cite{Gibbons} on the thermodynamics of de Sitter spacetime. They show that an observer in a de Sitter background experience a thermal bath of temperature $H/2\pi$, which they have interpreted as a thermal radiation emitted from the observer event horizon. The absortion of this radiation by the observer detector leads to a shrinking in the horizon area, which is interpreted as resulting from a decrease in the entropy beyond the horizon. Nevertheless, that original analysis does not consider the backreaction associated to the particle creation. As discussed above, the corresponding energy density and pressure obey, owing to the symmetry of de Sitter space, the equation of state $p_{\Lambda} = - \Lambda$, where $\Lambda$ is the effective cosmological term which defines the de Sitter horizon. Leading this into the conservation equation for the total energy (\ref{conservation}), we obtain equation (\ref{noconservation}). As already said, it shows that the particle creation (that is, its absortion from the thermal bath) is only possible if $\Lambda$ decays with time, with a consequent increase in the horizon area. This may be understood on the light of a reinterpretation of the Gibbons-Hawking thermal bath as resulting from the vacuum quantum fluctuations which originate the cosmological term. The horizon entropy gives the number of degrees of freedom (inside the horizon) associated to those fluctuations, and its increase just means that the extraction of particles from vacuum is an irreversible process. This interpretation is in better accordance with modern formulations of the holographic principle, as we will see below.

In this paper we shall explore the effect of the particle creation backreaction on the solution (\ref{H4}) presented above. We will show that it leads to an inflation solution with a scale invariant primordial spectrum. Furthermore, by assuming that inflation lasts for approximately $60$ e-folds, we will be able to obtain the scalar spectral index currently observed, $n \approx 0.97$.  However, before doing that, let us consider in the next section the role of particle creation in the opposite, low-energy limit of late times. We will see that it may also be responsible for the present cosmic acceleration, giving origin to a decaying vacuum term as well.

\section{The late-time limit}

In the case we consider the vacuum energy of interacting fields, it has been suggested that in a low energy, approximately de Sitter background the vacuum condensate originated from the QCD phase transition leads to $\Lambda \approx m^3 H$, where $m \approx 150$ MeV is the energy scale of the transition \cite{QCD}. This result is in fact intuitive. In a de Sitter background the energy per observable degree of freedom is given by the temperature of the horizon, $E \approx H$. For a massless free field this energy is disbributed in a volume $1/H^{3}$, leading to a density  $\Lambda \approx H^4$, as above. For a strongly interacting field in a low energy space-time, on the other hand, the occupied volume is $1/m^3$, owing to confinement, and the expected density is $\Lambda \approx m^3 H$. 

Such a late-time variation law for the vacuum term can also be derived as a backreaction of non-relativistic, light dark particles production in the expanding spacetime. Indeed, vacuum fluctuations of a particle of mass $M<H$ have a lifetime $\tau \approx 1/M$, longer than the Hubble time, being then observed as real particles irreversibly produced at a constant rate $\Gamma \approx 1/\tau \approx M$. We can also describe this process by saying that, as the wavelength of the fluctuations is greater than the horizon radius $1/H$, one of the particles in the pair scapes to the zone beyond the horizon, while the other appears as a created particle. If $M \lesssim H \ll 1$, the produced particles are non-relativistic (photons are not produced in the low-energy regime, because there exist no one-loop photon fluctuations), and the Boltzmann equation for this process is
\begin{equation} \label{Boltzman}
\frac{1}{a^3}\frac{d}{dt}\left(a^3n\right)= \Gamma n,
\end{equation}
where $n$ is the particle number density. By taking $\rho_m = nM$, it can also be written as
\begin{equation} \label{conservacao1}
\dot{\rho}_m + 3H\rho_m = \Gamma \rho_m.
\end{equation}

The particle production necessarily involves a backreaction. Let us take, in addition to (\ref{conservacao1}), the Friedmann equation
\begin{equation} \label{Friedmann}
\rho_m + \Lambda = 3H^2.
\end{equation}
If our background is approximately de Sitter, the vacuum pressure is $p_{\Lambda} = - \Lambda$. Using (\ref{conservacao1}) and (\ref{Friedmann}) we obtain the conservation equation for the total energy,
\begin{equation}
\dot{\rho} + 3H(\rho +p) = 0,
\end{equation}
provided we take 
\begin{equation}
\Lambda = 2\Gamma H + \lambda_0,
\end{equation}
 where $\lambda_0$ is a constant of integration. Since there is no natural scale for this constant, let us make it zero. Then we have $\Lambda = 2\Gamma H$. This is the time-variation law predicted for the vacuum density of the QCD condensate, with $\Gamma \approx M \approx m^3$. Dividing it by $3H^2$, we obtain
\begin{equation} \label{winfried}
\Gamma = \frac{3}{2} \left(1-\Omega_m \right) H,
\end{equation}
where $\Omega_m = 1 - \Omega_{\Lambda} \equiv \rho_m/(3H^2)$ is the relative matter density (for simplicity, we are considering only the spatially flat case). In the de Sitter limit ($\Omega_m = 0$), we have $\Gamma = 3H/2$, that is, the creation rate is equal (apart from a numerical factor) to the thermal bath temperature predicted by Gibbons and Hawking. It also means that the scale of the future de Sitter horizon is determined, through $\Gamma$, by the energy scale of the QCD phase transition, the last cosmological transition we have.

For the present time we have from (\ref{winfried}) (with $\Omega_m \approx 1/3$) $H_0 \approx \Gamma \approx m^3$, and hence $\Lambda \approx m^6$, where $H_0$ is the current Hubble parameter. The former result is an expression of the Eddington-Dirac large number coincidence \cite{Mena}. The later - also known as Zeldovich's relation \cite{Bjorken} - gives the correct order of magnitude for $\Lambda$. We also have $M \approx \Gamma \approx \sqrt{\Lambda} \sim 10^{-33}$eV. Note that this is the quantum of mass expected on the basis of the holography conjecture applied to de Sitter spacetime. According to the holographic principle \cite{Mena,Bousso}, the number of states inside the de Sitter horizon is given by the horizon area, $N \approx 1/H^2 = 3/\Lambda$. On the other hand, the energy inside the horizon is $E \approx \rho/H^3 = 3/H$. Therefore, the energy per degree of freedom is $E/N \approx \sqrt{\Lambda} \approx M$. As we have seen, in the de Sitter limit this quantum of energy is equal to the horizont temperature. By the way, let us remark that this tiny value for $M$ suggests that dark particles may be associated to pseudo-Goldstone bosons of the QCD condensate, with a dynamically generated mass given by the inverse of the scale of the de Sitter horizon. (The cosmological role of an ultra-light pseudo Nambu-Goldstone boson with mass $\sim H_0$ was already considered before, in another context \cite{Frieman}).

The corresponding cosmological model has a simple analytical solution, which reduces to the CDM model for early times and to a de Sitter universe for $t \rightarrow \infty$ \cite{Humberto}. It has the same free parameters of the standard model and presents good concordance when tested against type Ia supernovas, baryonic acoustic oscillations, the position of the first peak of CMB and the matter power spectrum \cite{Jailson3,Julio,Zimdahl}. Furthermore, as discussed in \cite{Julio}, the coincidence problem is alleviated, because the matter density contrast is suppressed in the asymptotic future, owing to the matter production. In the right panel of Fig. 2 we show the concordance in the $H_0 \times \Omega_{m0}$ space for those four observations. We consider the SDSS compilation of supernovas Ia calibrated with the MLCS2k2 fitter, since it is less dependent on a fiducial $\Lambda$CDM model than other supernovae samples. For the large scale structure (LSS) distribution we use the 2dFGRS data. We obtain for the present Hubble parameter $H_0 \approx 64$ km/s-Mpc, and marginalizing over it does not alter the matter density concordance value, $\Omega_{m0} \approx 0.45$. The best-fit universe age is $t_0 \approx 14$ Gyr. These results do not change if we use for supernovas Ia the Constitution sample compilated with MLCS2k2 \cite{Jailson3}. 

For the sake of comparison we show in the left panel the concordance region (with LSS excluded) for the spatially flat $\Lambda$CDM model with the SDSS (MLCS2k2) supernovae sample. The best-fit Hubble parameter is the same as in the $\Lambda$(t) model. However, as already pointed out in the literature, in this case there is a tension between the SDSS best-fit value for the matter density ($\Omega_{m0} \approx 0.40$) and the value obtained from the LSS analysis ($\Omega_{m0} \approx 0.23$) \cite{Jailson3}.

\begin{center}
\begin{figure}[t!]
\begin{minipage}[t]{0.375\linewidth}
\includegraphics[width=\linewidth]{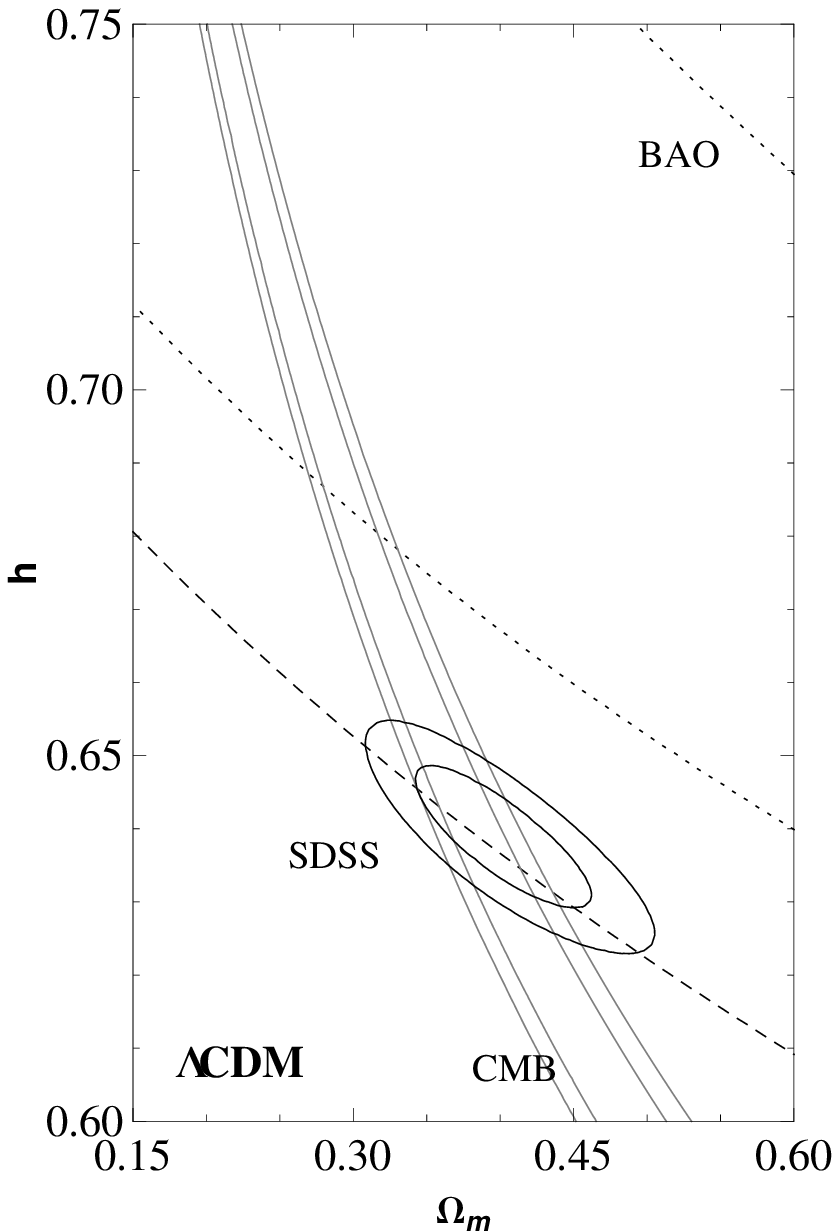}
\end{minipage}
\begin{minipage}[t]{0.5\linewidth}
\includegraphics[width=\linewidth]{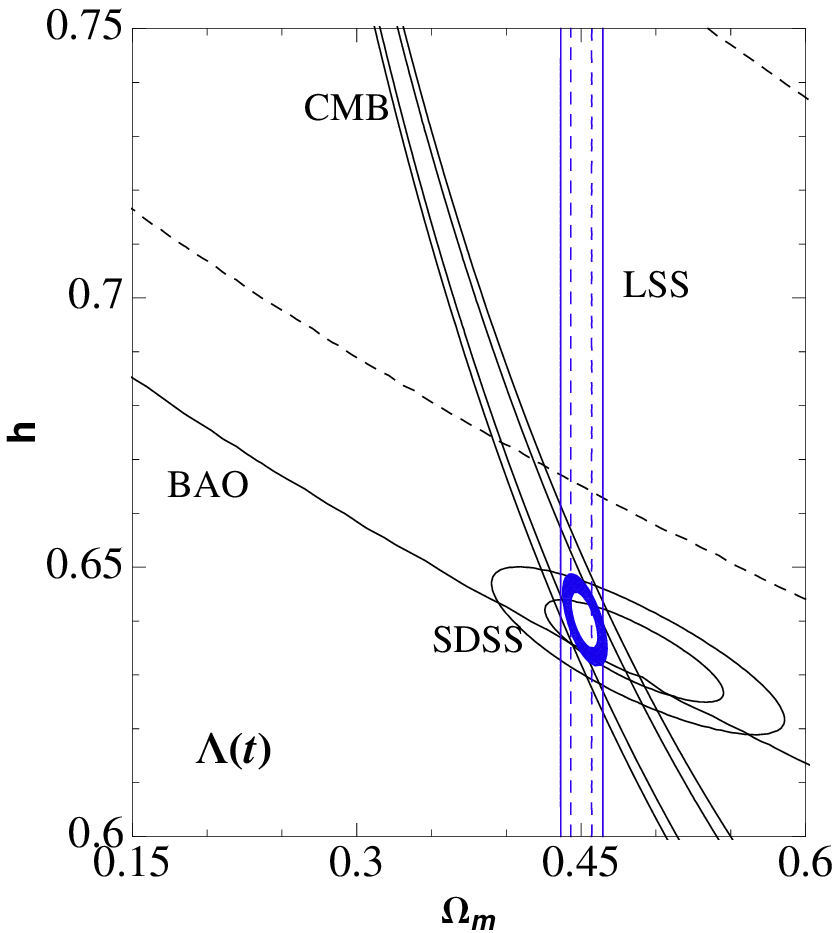}
\end{minipage}
\caption{{\protect\footnotesize {\bf Right.} Superposition of the four tests (SNe Ia, BAO, CMB and LSS) performed with the present model. The blue ellipse corresponds to the $2\sigma$ concordance region. {\bf Left.} SNe Ia, BAO and CMB $2\sigma$ confidence regions for the spatially flat $\Lambda$CDM model \cite{Jailson3}.}}
\end{figure}
\end{center}

\section{The inflation solution}

Let us return to the early-time limit of high energy densities. In this case the particles are relativistic, with an average energy $\mu \approx H$. That this is the typical energy of any degree of freedom, including vacuum fluctuations, can be seen by applying again the holographic principle \cite{Mena,Bousso} to a de Sitter (or quasi-de Sitter) spacetime. The number of degrees of freedom inside the de Sitter horizon is proportional to its surface, $N \approx 1/H^2$, while the total energy inside the horizon is $E \approx \rho/H^3 \approx 1/H$. Therefore, the energy per degree of freedom is $\mu = E/N \approx H$. Another way to support this result is remembering that a temperature $H/2\pi$ can be associated to the de Sitter background.

The lifetime of a vacuum fluctuation is now given by $\tau \approx 1/H$. If it is a bit longer than $1/H$, we have relativistic particles production at a rate $\Gamma \approx H$. The Boltzmann equation has now the form
\begin{equation} \label{Boltzmann}
\frac{1}{a^4}\frac{d}{dt}\left(a^4n\right)= \gamma H n,
\end{equation}
where $n$ is the particle number density and $\gamma$ is a constant of the order of unity. In an approximately de Sitter spacetime (where $\mu \approx$ constant), (\ref{Boltzmann}) can be rewritten as
\begin{equation} \label{conservacao}
\dot{\rho}_m + 3H(\rho_m  + p_m)=\gamma H \rho_m,
\end{equation}
where $p_m = \rho_m/3$. Equations (\ref{conservacao}) and (\ref{conservacao1}) are characteristic of models with interaction in the dark sector \cite{Chimento,delCampo}.

Again, the particle production involves a backreaction. From Eq. (\ref{conservacao}) (with $p_{\Lambda} = -\Lambda$)  and the Friedmann equation
\begin{equation} \label{Friedmann2}
\rho_m + \Lambda = 3H^2,
\end{equation}
we derive the conservation of the total energy,
\begin{equation}
\dot{\rho} + 3H(\rho +p) = 0,
\end{equation}
provided we take
\begin{equation} \label{Lambda}
\Lambda = \frac{3\gamma}{4} H^2 + \Lambda_0.
\end{equation}
Since there is no natural scale for the integration constant $\Lambda_0$, we take it equals to zero. Let us remark that a quadratic dependence of $\Lambda$ on $H$ was already considered in other contexts. For example, it can be obtained on the basis of renormalization group results \cite{Sola}.

The solution for (\ref{conservacao}), (\ref{Friedmann2}) and (\ref{Lambda}) is given by
\begin{equation} \label{H}
\frac{1}{H} = \epsilon t + \frac{1}{H_0},
\end{equation}
where $\epsilon = (4-\gamma)/2$ and $H_0$ is now an integration constant. For $\gamma \approx 4$ we have $\epsilon \ll 1$. Hence, for $\epsilon|t|\ll1$ we have $H \approx H_0$, that is, an approximately de Sitter solution, as assumed above.

We can verify from (\ref{H}) that $\epsilon$ is the slow-roll inflation parameter defined by $\epsilon = -\dot{H}/H^2$. As it is constant, the second slow-roll parameter is $\delta = -\epsilon$, and the scalar spectral index is \cite{Dodelson}
\begin{equation}
n_s = 1 - 2\epsilon.
\end{equation}

From (\ref{H}) we define the time when inflation ends by $H_0 t = 1/\epsilon$. Therefore, assuming that inflation lasts for about $60$ e-folds, as required to solve the horizon problem, we have $\epsilon \approx 1/60$, leading to $n_s \approx 0.97$. This is approximately the current value obtained for the scalar spectral index from CMB observations, $n_s = 0.96 \pm 0.01$ ($1\sigma$) \cite{CMB}. Note that such a relation between the number of e-folds and $n_s$ is not general, holding only for models with constant $\epsilon$. On the other hand, it is easy to show from (\ref{Friedmann2}) and (\ref{Lambda}) that the matter relative density is $\Omega_m = \epsilon/2 \approx 0.01$. That is, during inflation radiation and matter represented around $1\%$ of the energy content.

Up to now we have been considering $\gamma$ as a constant in (\ref{Boltzmann}). However, it actually decreases as long as $H$ decreases, because the production of a particle species stops when $H$ crosses the threshold of the particle mass\footnote{Strictly speaking, the temperature and entropy associated to the horizon are not well defined when the spacetime departures significantly from de Sitter.}. Eventually $\gamma$ goes to zero, hence we have $\epsilon = 2$ and, for $\epsilon t \gg 1/H_0$, it follows from (\ref{H}) that $Ht = 1/2$, which characterises a radiation phase, with $\Omega_m = 1$. As for the ultra-light dark particles considered above, their production starts after the QCD phase transition, but they are subdominant during the radiation phase \cite{Humberto}.

The energy scale of inflation (given by the integration constant $H_0$ in (\ref{H})) is not determined in this model. It is reasonable to assume that it is above the electroweak scale, in which case all the observed particles are produced. The Planck scale ($H \approx 1$) is a natural superior limit, at which the ansatz $\Lambda \approx H^4$ discussed at the beginning is valid. In this case inflation would take place at the end of the non-singular quasi-de Sitter phase.

As in this model we have two interacting fluids, namely the vacuum term and relativistic matter, the presence of entropic perturbations must be investigated. Supposing that both components have no intrinsic non-adiabatic perturbations, the only entropic contribution is given by \cite{Zimdahl}
\begin{equation}
\hat{p}- \frac{\dot{p}}{\dot{\rho}}\hat{\rho} = \frac{\dot{\Lambda}\dot{\rho}_{m}}{\dot{\rho}}
\left(\frac{\hat{\rho}_{m}}{\dot{\rho}_{m}} - \frac{\hat{\Lambda}}{\dot{\Lambda}}\right),\label{nadtot1}
\end{equation} 
where a hat means perturbation of a given quantity.

Owing to its equation of state, $p_{\Lambda} = -\Lambda$, the vacuum component perturbation is negligible for scales inside the horizon \cite{Zimdahl}. Therefore, the relative entropic perturbation (\ref{nadtot1}) reduces to
\begin{equation}
\hat{p}- \frac{\dot{p}}{\dot{\rho}}\hat{\rho} \approx \frac{\dot{\Lambda}}{\dot{\rho}}\hat{\rho}_{m} \approx \hat{\rho}_m,
\end{equation} 
where in the last approximation we have used the fact that, during the inflationary phase, $\Lambda \approx \rho$. The non-adiabatic perturbations are then proportional to $\hat{\rho}_m$, and  their presence does not affect the scale-invariance of the primordial espectrum.

In reference \cite{Tavakol}, both the Planck limit $\Lambda \approx H^4$ and the low-energy result $\Lambda \approx MH$ were also modeled, alternatively, by a self-interacting scalar field, with the interaction potential $V(\phi)$ playing the role of a decaying vacuum term, and the kinetic term $\dot{\phi}^2/2$ representing produced stiff matter. Let us show that we can do the same here, that is, we can model this inflationary scenario by introducing an equivalent inflaton field. The field equations are \cite{Tavakol}
\begin{equation}
3H^2 = V + 2H'^2,
\end{equation}
\begin{equation}
\dot{\phi} = -2H',
\end{equation}
where the prime means derivative with respect to $\phi$. From equation (\ref{Lambda}) we see that, in this case,  $V = 3\gamma H^2/4$. The solution is then given by
\begin{equation}
\frac{1}{H} = \epsilon t+\frac{1}{H_0},
\end{equation}
\begin{equation} \label{potential}
V(\phi) = \frac{3\gamma H_0^2}{4}\, e^{\pm \sqrt{2\epsilon}\phi},
\end{equation}
where now $\epsilon = (3/2) (4-\gamma)/2$. The first equation is, apart a re-scaling of time, the same solution (\ref{H}) obtained before. The second equation gives, for $\epsilon \ll 1$ (that is, $\gamma \approx 4$), an approximately constant potential $V \approx 3H^2$.
It is easy to verify that $V''/V = (V'/V)^2 = 2\epsilon \ll 1$. This guarantees that we have a slow-roll inflation, with a scale-invariant spectrum $k^3P = H^2/9\epsilon$ \cite{Dodelson}. It also means that, as asserted above, the perturbations in the potential (i.e., in the vacuum term) are negligible, since $\hat{V} = 3H^2 \frac{V'}{V} \hat{\phi} \ll \hat{\phi}$. The potential (\ref{potential}) was also considered by \cite{winfried} in the context of a model with particle creation.

\section{Conclusion}

Particle creation is something expected in expanding spacetimes \cite{Parker}. In spite of the difficulty in deriving the production rate and backreaction in general, this phenomenon may in principle be related with inflation and the present cosmic acceleration, a possibility already considered in different ways by some authors \cite{ademir, winfried2} . We have shown in this paper that the production of non-relativistic, ultra-light dark particles at late times leads to a concordance model in which the vacuum density decays with time. In the limit of high energies, on the other hand, we have shown that the production of relativistic particles from the vacuum leads to a viable inflationary solution.


\section*{Acknowledgements}

I would like to thank the International Centre for Theoretical Physics (Trieste, Italy) for the hospitality during the completion of this paper. This work is partially supported by the Brazilian Council for Scientific Research (CNPq), grants \# 305133/2008-0 and \# 472341/2009-0.


{}

\end{document}